# The Role of Digital Agriculture in Transforming Rural Areas into Smart Villages

Mohammad Raziuddin Chowdhury[1], Md Sakib Ullah Sourav[2], Rejwan Bin Sulaiman[3]

[1] American International University, Bangladesh
[2] STEM in Bangla, Chattogram, Bangladesh
[3] University of Bedfordshire, Luton, United Kingdom

Emails: razichy3@gmail.com[1], sakibsourav@outlook.com[2], rejwan.binsulaiman@gmail.com[3]

**Abstract**

From the perspective of any nation, rural areas generally present a comparable set of problems, such as a lack of proper health care, education, living conditions, wages, and market opportunities. Some nations have created and developed the concept of smart villages during the previous few decades, which effectively addresses these issues. The landscape of traditional agriculture has been radically altered by digital agriculture, which has also had a positive economic impact on farmers and those who live in rural regions by ensuring an increase in agricultural production. We explored current issues in rural areas, and the consequences of smart village applications, and then illustrate our concept of smart village from recent examples of how emerging digital agriculture trends contribute to improving agricultural production in this chapter.

## 1 Introduction

Technology has been making impressive headways toward creating a world free of poverty, hunger, and diseases, and the implementation of technological inventions has transformed human lives in ways that were unimaginable even a few decades ago. However, in this march towards perfection, a large number of people, specifically those living in impoverished rural areas all over the world, have been mostly left behind. In many underdeveloped and developing countries, rural villages are often impeded by insufficient access to the internet and electricity. According to EU Rural Review, a mere 47% of rural homes have internet access as opposed to more than 80% in urban areas [1]. In addition, more than 1.3 billion people are deprived of access to electricity all over the world [2]. This leads to lacking healthcare, transportation, and education system, extreme poverty, inadequate digital literacy and digital gap, migratory imbalance, and a poorly structured economy [3]. The situation has been referred to as a 'vicious circle driving rural decline' [4]. Inclusive digital transformation through a regionally adaptive, resident-oriented, and knowledge-based policy, focusing on sustainable growth is vital to tackle the issue and to ensure a better quality of lifestyle for everyone [34]. The potential of the smart village concept in this regard is therefore being recognized by world leaders along with common agricultural policy (CAP) remodeling in progress [4].

It is now well understood that rurality is both a demographic or material construction and a symbolic or imaginary one [78, 79]. It is not easy to think about this in a global context other than to say that there is a tendency to position rurality as a receding and even vestigial space in the face of the big story of modernity, which is positioned as urban. However, in reality, rural areas have to encounter several challenges. Such as, healthcare, education, marketplace, and so on.

Rural health refers to the well-being of residents of rural areas, who are typically more remote from medical facilities and other services than their urban counterparts. Rural communities experience far worse healthcare services than urban populations, including difficulty with mental illness, substance abuse, general health, and sexual health. These discrepancies in access to healthcare can be observed all over the world. If we look at statistics, 22% of urban people lack access to healthcare in contrast to 56% of rural residents which gives us a concrete idea of how compelling the situation is for rural health [5]. As young people relocate to urban centers for education and work, these places frequently encounter a population decrease. Aging rural populations are the outcome, and they have greater health demands and less capacity for communal activities. The life expectancy and health conditions of people who live in rural and distant areas are lower than those in urban areas. Rural health disparities refer to these types of variations in risk. Rural residents experience social exclusion as a result of socioeconomic disparities, poverty, and lower employment rates, as well as disparities relating to services, resources, and transportation. All of these factors also influence the quality and quantity of healthcare personnel who are willing to provide services in faraway areas. This number will continue to decline in the future and has already become a core problem in delivering primary healthcare which is also a headache for healthcare policymakers around the globe. People in rural locations find it particularly challenging to receive healthcare and maintain good health because of this. Countries encounter significant difficulties in service delivery, human resources, governance, financing, communication, and in some areas, corruption when formulating a rural health policy [35, 36]. In every nation, rural residents' health is in poorer shape than that of their urban counterparts.

Over the years, a persistent cycle of unfavorable, self-fulfilling beliefs regarding rural education solidified itself in the minds of policymakers. This low expectation resulted in insufficient effort and resources, which in turn made subpar standards normal. When rural high school graduates try to join the workforce, their prospects are not always tightly linked to urban economies as those of their urban counterparts. The major reasons behind such a trend are the unavailability, and lack of awareness regarding resources available online for free among village people. This is especially true for third-world countries. On the other hand, in advanced capitalist countries, the idea of public education has developed alongside the phenomena of urbanization. As a result, there is an assumed and seemingly natural relationship between the growth of education and urbanization. In fact, the notions of urbanization, bureaucratization, and education are often used as a stand-in for modernity [37]. Rural schooling is also impacted by the prolonged problem of rural poverty.

A rural economy can contribute significantly to producing employment, fostering economic growth, and fostering sustainable development. Among the extremely poor population of 1.2 billion around the globe, 75% reside in rural areas. This statistic shows the potential for improvement as well as negligence toward the rural economy. The impoverished in rural areas face several obstacles while trying to compete as customers, producers, or business owners. Routine commercial interactions are more challenging and expensive for businesses in rural locations due to geographic isolation. Rural residents may even find it more challenging to obtain government funding possibilities due to the frequent lack of professional staff needed to compose and submit competitive grant applications, especially in LIMCs (Low- or Middle-Income Countries). Lack of diversity in the rural economy, suggests that agriculture dominates as the primary form of livelihood in rural surroundings. This can be used to our advantage. Concentrated efforts in agriculture in the form of structure, investments, and training can result in significant improvements in the rural economy.

The rest of the parts of this chapter is aligned as follows, section 2 describes previous models of smart villages from different perspective proposed by the researchers. In section 3, we demonstrate our proposed model of a smart village in view of digital agriculture applications in four specific areas. Finally, section 4 concludes this chapter.

**2 Smart Village Models in Recent Times**

According to European Network for Rural Development (ENRD) -"Smart villages are communities in rural areas that use innovative solutions to improve their resilience, building on local strength and opportunities. They rely on a participatory approach to develop and implement their strategy to improve their economic, social, and/or environmental condition, in particular by mobilizing solutions offered by digital technologies" [6]. The primary objective is, therefore, to capitalize on locally accessible resources and opportunities and generate cost-effective schemes to overcome individual community-based hurdles, and strive toward building an autonomous and empowered community [7]. Participation in rural communities, multi-stakeholder engagement, and improving connectivity are considered key features of this concept [8].

Although technological transformations are frequently considered a fundamental element of smart villages, they are often not the most crucial and definitely not the sole component [9]. Director of the European Network of Broadband Competence Offices Support Facility Jan Dröge states that- "Smart Villages are not exclusively about digital services, but digital transformation can be an important element of rural development and the regeneration of rural areas." Digitalization is a tool that assists in the implementation of the overall concept, it is not, however, the core purpose. The comprehensive policy context coincides with other aspects of sustainable development goals (SDGs), endorsed by the United Nations in 2015, aiming to eradicate poverty, hunger, and inequality by 2030. For instance, Prophet Elijah's Hospice Foundation in Michalow, Poland came up with a smart modus operandi to deliver personalized health care to the chronically ill senior citizens of the areas nearby. Doctors, nurses, and professional caregivers visit the patients as often as required, keep them company, and provide training and emotional support to the families of the sick so that they can take better care of the elderly at home. Their service does not make use of advanced telecommunication, but it still manages to deliver a smart solution for a social concern [10]. Another Swedish village called Vuollerim adopted crowdsourcing as a means of transforming their community into a smart one. The idea entails employing all available community resources to support innovative projects toward local development and the growth of the local economy [11].

The Smart Villages strategy delves into a vast range of policies and there can be no one-size-fits-all approach that can fit the context of each community and cater to their unique circumstances. There is no single route to being smart [12]. For instance, the Millennium Village Project (MVP) was a high-profile, cross-sectoral rural development program, which was launched across 10 different sub-Saharan African villages in 2005, with a view to attaining the Millennium Development Goals (MDGs) by 2015. Despite the tremendous financial backing from the UN and other political and non-political organizations, it failed to deliver on its promises in Sauri village in western Kenya. From a study conducted by the Department of Social Sciences of Wageningen University in the Netherlands, it was concluded that MVP fell short of expectations in Sauri Millennium Village due to the lack of consideration given to the diversity and uniqueness of different social agencies, social challenges, and necessities across the operating regions. In the agricultural intervention scheme, MVP failed to address the general tendency of Sauri villagers to reject modern inventions and prevalent market trends. Consequently, it brought about conflicts and mistrust among the people owing to the appropriation of agricultural resources by the privileged, worsening of the prevailing inequalities, community-wide corruption, and injured social relations [13]. Whereas, the overall assessment of MVP exhibits significant positive impacts on one-third of its rural development goals including considerable agricultural and health system benefits [14].

Smart village is a fairly recent concept and the idea of it first surfaced globally as a response to the Envision2030 project approved by the UN in 2015. It contained 17 sustainable development goals to fight imbalances in the economy, poor infrastructure, lack of connectivity, global warming, and so on. The idea also came up in the writings of Gary Paul Green, Terry van Gevelt and John Holmes in which they put forward a proposition for rural development [15,16]. The smart village concept acquired momentum in the EU in 2016, at the CORK 2.0 European Conference on Rural Development. Fostering rural evolution, bolstering rural economy and entrepreneurship, abetting rural viability and resilience, conservation of rural atmosphere, preservation of natural deposits, developing climate consciousness, advocating knowledge and creativity, reinforcement of rural administration, effective implementation, and elucidation of policies, and monitoring of performance and developing a sense of liability were announced as the base for a thorough and integrated rural and agricultural policy in CORK 2.0 [17]. It received further stimulus in 2017 after being initiated by the European Parliament and the publication of 'EU Action for Smart Villages' by the European Commission in collaboration with the European Parliament [9]. A major contribution of this initiative involves the formation of a thematic group consisting of leaders and professionals from all over Europe to create a model blueprint for the boosting and furtherance of the Smart Village concept. Common Agricultural Policy (CAP), one of the most influential rural improvement policies, is financed jointly by European Agricultural Fund for Rural Development (EAFRD) and European Agricultural Guarantee Fund (EAGF) [7]. In October 2017, 250 different government officials, spokespersons of several organizations in the EU, and civil society groups from 40 European nations gathered at the Dutch village called Venhorst to talk about current pressing issues in Europe and the potential the rural communities possess in dealing with these concerns [18]. The Commission's pledge to augment support for rural administrations and communities that aim to establish smart villages was reaffirmed in its statement on the Future of Food and Farming, which was released in November 2017 [19]. Further, the Bled declaration on 13[th] April 2018 stated that- "the rural digital economy, if developed in an innovative, integrated and inclusive way, has the potential to improve the life-quality of rural citizens and, thereby, contribute to tackling the current depopulation of- and the migration from- rural areas" [20]. It recommended a number of measures such as digitization of basic civic services, exploitation of inexhaustible sources of energy, minimizing waste generation, site-specific crop management, creation of high-paying employment opportunities through rural tourism, shared and bio-based economy, and social innovation in order to accomplish the target of converting rural areas into smart villages in some EU Member States by 2019.

As the concept is still in its infancy, the implications of the idea of a Smart Village still vary depending on the locality and their perspectives. While researchers in Europe focus their attention on technological advances and effective public infrastructure, Asia, America, and Africa are working on building climate-conscious, energy-sufficient, and agriculture-based communities [21]. There are several fascinating instances of smart villages across the world. For instance, as part of the Save and Grow project run by the UN's Food and Agricultural

Organization (FAO), smallholder farmers in six villages in Anuradhapura, Sri Lanka have been trained in climate-smart agricultural (CSA) techniques [22]. This project also provided a foundation for future evaluations by supplying data on the project outcomes. As defined by the World Bank, "CSA is an integrated approach to managing landscapes—cropland, livestock, forests, and fisheries--that address the interlinked challenges of food security and climate change [23]." Climate-related agricultural predicaments in Sri Lanka include torrential precipitation, sedimentation in lowland water reservoirs and soil degradation in upland areas, etc. All these, paired with the absence of technological and scientific knowledge and monetary assistance negatively impact crop yields, food security, and the overall economy. However, through the application of efficient water management practices e.g. early plantation, proper utilization of rainwater and alternate wetting and drying technique, and the use of scientific equipment such as soil testing kits and leaf color charts, farmers were able to trim down water requirements, increase water availability and cut down the use of fertilizers [24]. Several other CSA projects in different parts of the world including Egypt, Somalia, Mongolia, Cambodia, Georgia, Kyrgyzstan, Moldova, Senegal, Mali, Ghana, Botswana, Ecuador, Saint Lucia, Switzerland, Italy, and the African Region are being undertaken.

In South Korea, rural tourism was adopted to deal with the issues of existing economic and social disparity between rural and urban areas in the early 1990s. It was anticipated that Korean farmers would be able to benefit from the diversified tourism sector and get a better return from agricultural produce which in turn will lead to the revitalization of the rural economy. Thus the Korean government put enormous effort to make this new initiative successful and in 2003, disclosed plans to spend around 119 trillion Korean won on rural and agricultural development from 2004 to 2013. Another 7 billion won was granted to fund the construction of rural development zones in cities with rural features [25]. However, the project was not able to reach its full potential due to discriminated distribution of government funding, the absence of appropriate managerial programs, and the failure of the programs to draw in the urban populace. Rural tourism has also been embraced by other countries throughout the world including Indonesia [26], Romania [27], China [28], and so on.

Italy has introduced one of Europe's most well-rounded measures to deal with the challenges of emigration and limited services accessibility titled the National Strategy for Inner Areas for the 2014–2020 programming timeframe. 23% of Italians reside in Inner rural areas which are distinguished by their detachment from the major commercial and service hubs. The plan of action is built on a community-based, multi-fund strategy that operates under multidimensional leadership and coordination among different administrative entities intending to achieve concurrent improvement in a number of service areas. Some of the initiatives that have received assistance to date include - a communal carpooling program run by a local cooperative through an online platform in Val Maira, Piedmont; cyber schooling at high schools in Piacenza-Parma Apennine (Emilia Romagna) and Beigua Sol (Liguria); utilization of remote diagnostics technologies in Matese's (Molise) neighborhood drug-stores; provisioning advanced equipment for greater protection against landslides in Madonie (Sicily) [29].

The prime minister's office in Finland issues open invitations on a yearly basis for research proposal submissions aligning with the major goals of the government, such as - employment and competitiveness, knowledge and education, health and wellbeing, digitization, experimentation, deregulation, and reforms [30]. The major barrier to the digitalization of rural regions in Finland, according to the 2016 Smart Countryside side research performed by the Finnish ministry of transport and communication, is the lack of digital competence and interest among a huge percentage of the rural people. From the study, it was concluded that aid for working internet service and digital advisory services based on local demands and expertise, and innovative digital solutions to local challenges are vital to minimize the likelihood of digital exclusion. It also emphasized that the financial advantages of such measures should be widely publicized so that people may perceive them. The government's commitment to supporting rural digitization, the Strategy for Digital Infrastructure, and the Smartest Village 2020 competition are just a few of the responses the study was able to elicit [30].

Since time immemorial, agriculture has played a crucial role in rural economies and sustainability. The agricultural industry of the EU, which is tightly interwoven into the worldwide economy accounts for 3.7% of their total GDP, ensures food for nearly five hundred million people, and creates job opportunities for another 44 million [19]. Consequently, it is considered to be one of the most significant tools in the rural development model. The long-term viability of smart communities depends on the innovation and implementation of new and improved methods. Hence, precision agriculture utilizes cutting-edge technologies for raising animals and crops that not only save time but also eliminate waste and satisfy the standards of smart communities. The history of agriculture is marked by three remarkable technological shifts beginning with a labor-intensive system that yielded a relatively low return. That all changed as the second industrial revolution became more widespread around the turn of the $20^{th}$ century. This is referred to as Agriculture 2.0. Later in the 1960s, a significant boost in output was achieved through the use of pesticides and chemical fertilizers. Currently, the world is witnessing another technological revolution and advancing towards Agriculture 4.0 [31]. However, over the last ten years, the agricultural sector has been struggling as is evident by the dwindling number of farmland, growing level of indigence among farmers, exponentially rising prices of food, and so on. This can be attributed to a number of different issues, namely-weather hazards, water scarcity, diseases, unpredictable market dynamics, resource constraints, etc. Additionally, given the urgent nature of climate change, it is possible that things may get worse. Furthermore, the apparent lack of digital literacy and elderly farmers' inability to participate in digitization severely limit the possibilities for progress. Rural farmers are often deprived of access to the vast amount of information and services that the internet has to offer regarding farming practices and market policies due to inadequate connectivity. Therefore, a feasible agricultural intervention has become a necessity. Thus it is safe to anticipate technology to have a much stronger influence on the agri-food industries in the years ahead.

According to the International Society of Precision Agriculture (ISPA) – "Precision Agriculture is a management strategy that gathers, processes and analyzes temporal, spatial and individual data and combines it with other information to support management decisions according to estimated variability for improved resource use efficiency, productivity, quality, profitability and sustainability of agricultural production [32]." Thus, to ensure optimization in resource utilization and abatement of potential environmental pollution, the farmers tailor their course of action to the precise requirements of the livestock and vegetation. It incorporates bio and nanotechnologies, Internet of Things (IoT) and blockchain-based procedures, drone technologies, and other climate-smart strategies that might offer remedies to the aforementioned issues [33]. To establish precision agriculture as a primary element of the smart village movement, it is customary to find ways to fight the obstacles such as insufficient internet access, poor digital infrastructure, deficiencies in digital skill development and digital literacy, etc. that are holding back the uptake of these technologies.

## 3 Proposed Model of Smart Village

Apart from many prior ideas of smart village, we feel to denote the term "smart village" in light of digital agriculture applications in modern days (Figure 1). The use of state-of-the-art equipment, and automation alongside information and communication technologies (ICTs) to enhance decision-making and production has been altering agriculture as a major part of the digital revolution. Increased yields, cost reduction, and environmental damage reduction on crops are possible through the usage of GPS, remote sensing, IoT, AI, and machine learning. Technology has been revolutionizing agriculture and enabling farmers to get maximum output. Ideas such as precision agriculture that involve extensive data collection and analysis regarding crop-specific factors are contributing to improving crop yield. In this study we will summarize the contribution of digital agriculture in four main categories:

- Precise Irrigation
- Detecting and controlling crop diseases and pests
- Soil fertility analysis
- Weed management

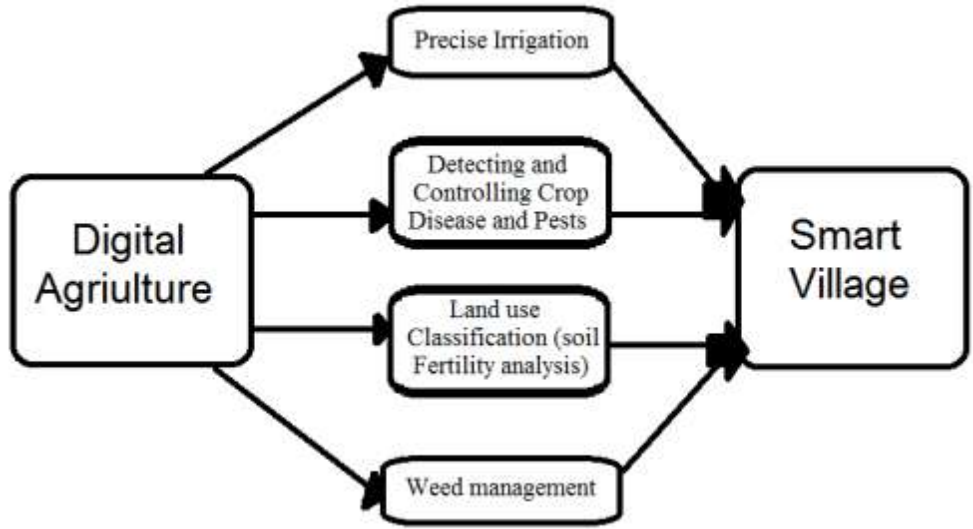

**Figure 1**. Methodology of our proposed smart village with digital agriculture applications in four areas.

### 3.1 Precise Irrigation

Irrigation is the process of supplying water to lands to foster the growth of crops/ plants through man-made measures such as canals, sprinklers, etc. In regions with intermittent rainfall, irrigation is essential for food production. For the best yield of crops, the irrigation system needs diligent monitoring. This takes up a lot of time, labor, and cost. The amount of water flow required depends on a wide range of factors including soil moisture, temperature, crop health, etc., meaning water requirement for crops is not constant and changes depending on seasons and crop growth. While farmers around the globe, especially in underdeveloped countries, use age-old techniques in accordance with their invaluable experience, digital agriculture has started improving the ease of operation. With data-driven decisions regarding water flow management and the inclusion of energy-efficient methods in remote areas have helped to reshape traditional irrigation systems. Implementation of such smart irrigation systems also contributes massively to water conservation by controlling the supply of water depending on smart sensor data. Smart irrigation systems utilize a number of technologies, including IoT, cloud computing, wireless sensor networks, machine learning, artificial intelligence, smartphones, etc.

IoT or "Internet of Things" is an expanding network of interconnected objects or devices with integrated sensors that can gather and share data among themselves. The inclusion of IoT in the irrigation system has been slowly gathering momentum. Commercially manufactured sensors available for irrigation and farming systems are still quite costly to incorporate for scrambling farmers. But in recent times a lot of attention has been directed toward low-cost sensors. This has been a huge boost for IOT-based smart agricultural systems. Research on different applications of sensors to measure parameters such as plant water stress is being conducted. Alongside traditional ground soil moisture measurement sensors, new low-cost and low-power consumption technology such as leaf sensing is being evaluated [38]. Different proposed models of smart irrigation are taking up different approaches for data acquisition through different types of sensors, refining the technology further in the process. The advancement of these sensors is enabling precise measurement of data and in turn, making the decision-making depending on the data easier and more accurate. Most smart irrigation systems employ an automatic water supply system based on these data. Which removes humane inspection, and labor from the equation. IoT system with smartphone application provides farmers with accurate, real-time information about soil moisture, water level in reservoirs, and weather forecast data acquired through different sensors. These applications are often used to alert users regarding weather conditions in near future and suggest strategies to mitigate adversities. Farming operations are benefitting from such data-based insights provided by such applications. Over-irrigation and under-irrigation are problems that no amount of humane experience can solve. With a smart precision irrigation system's array of sensors, we can avoid both of these problems.

These IoT systems' improvement has also been reinforced by the improvement in wireless technology. Wifi is the most used form of communication technology due to its availability and low financial implications. Although a large area of coverage is a problem for this technology. In most cases for long-range communication between the array of sensors and the system, Global System for Mobile communication (GSM) is used. Another noticeable wireless technology is LoRa which has been mostly used in areas with no service over a long range of land [39]. The improvement in this sector means that IoT systems can be incorporated even in the most peripheral locations.

Irrigation takes up 70% of the world's freshwater usage [40]. Smart water management is hence imperative to ensure food and water security. Smart irrigation systems are built on the principle of providing the exact amount of water needed, which decreases the usage of excess water. Irrigation scheduling is one of the major smart agriculture concepts. It involves decision-making of when and how much water is to be provided. These decisions are taken based on specific factors such as water required by plants, development stage of the crop, condition of the soil around the root of the plants, etc. Various irrigation scheduling tools are working on different approaches depending on the crop/plant of discussion. Monitoring in smart irrigation can be divided into three major categories: soil, weather, and plant-based [41]. These tools can help reduce water and energy usage significantly when plant-based indicators are applied [42]. A typical third-world farmer uses his experience to decide on irrigation volume and timeline, without the help of any precise or conclusive data. This process includes a lot of manual experimentation and frequent checking of the several factors stated above. Irrigation scheduling tools help mitigate the error in human judgment. Thus integration of irrigation scheduling automation reduces the usage of excess water.

Irrigation systems also require a substantial amount of energy to operate. It is a major challenge, especially in remote areas where energy sources are scarce. Without stable power grid connections, farmers from such areas use fossil fuels to operate pumps. With the deficit in electricity in many countries and the high price of fuel, various alternative approaches have been explored. Solar energy-based water pumping has been looked into as a viable option. Operating costs can be trimmed down with the implementation of such projects. Apart from being cost-effective PV solar cells are also a form of renewable energy. Wind-powered pumps are another sustainable energy option to implement as a source of electrical energy for smart irrigation. A solar or wind power-based IoT system for irrigation is an option to be considered to replace the current water and energy exhaustive models in place.

Many different adaptations of smart irrigation have been adding to the evolution of the technology for the better. There are many different forms of farming such as tunnel farming, greenhouses, etc. An array of work has been carried out to make the technology better for such specific farming methods [43, 44] The architectures of different proposed models have many features unique to them. Garcia et al. proposed a four-layer model for the implementation of smart irrigation [45]. The recommended layers are devices, communication, services, and application. Many models implement a machine learning model to provide a final decision based on ontology and sensor data [46].

Many other elements of irrigation have been under scrutiny. Water sprinklers and canal lining are among the most prominent topics in such discussions. Although at fairly early stages of development, machine learning and artificial intelligence concepts are also being worked on, to further reduce human involvement in the process of irrigation. In the future, these AI-based systems may be used for making inferences about irrigation water amounts and schedules for specific plants or crops to get maximum yield with minimum/ optimum water and energy usage.

When evaluating performance in a specific sample space, compared to traditional irrigation models, intelligent irrigation scheduling systems have been able to reduce water usage and electricity consumption by 59.61% and 67.35% respectively, all the while increasing crop yield by 22.58% [47]. For optimum utilization of all available resources such as water, energy, labor, and time there is no replacement for digital irrigation systems. Ever increasing population and global climate change will have adverse effects on earth's natural resources in the coming days. So it is high time that utmost importance is given to technologies such as smart irrigation and precision agriculture to keep the water reserves safe.

**3.2 Detecting Pest and Crop Diseases**

Pest control is the process of eliminating species that are deemed detrimental to plant/ crop health and yield. One of the main issues regarding pest control is early-stage detection.

Traditionally farmers try to manually detect the presence of pest attacks. Which is an inefficient method to find such small anomalies over a vast land of crops. The tedious process of manual pest detection involves frequent monitoring and does not guarantee concrete results. The potential loss of crop yield to pests is about 18% [48]. In traditional agricultural systems pest attacks are often detected at the later stages of the spread. As a result, large amounts of pesticides are applied to manage agricultural damage. This pesticide usage results in soil damage, water body pollution, aquatic species death, and depletion. Also, excessive use of pesticides is often detrimental to the health of people consuming the food produced in the pesticide-dependent pest control land. Hence for sustainable agriculture, the adoption of better technology to detect pest attacks at the early stages of infection is imperative.

In recent years a lot of effort has been dedicated to diagnosing disease and pest attacks from afar and more accurately. Image processing techniques have been quite useful in such applications. Through the use of digital image processing, visual elements that are important to a specific application can be highlighted and details that are not as important for the application can be toned down. The processed image is then used to gather relevant features. Image processing has been used in accordance with various smart sensors, microcontrollers, and prediction algorithms to efficiently determine pest attacks.

In one such application of combining image processing and sensory data, Passive Infrared (PIR) Sensors and motion sensors are used to detect the presence of any pests based on heat signature and movement. After the detection of pests images are captured and processed to send to a microcontroller. Then acoustic sensors are used to gather sounds of the pests laying eggs and feeding. The collection of both image and audio data is to avoid any false alarms regarding an eco-friendly pest. These data are then compared with the database using a microcontroller to determine if the pest is potentially harmful. If the pests are detected to be harmful an ultrasonic generator is used to create sound (at a frequency inaudible to humans) that creates stress on the pests' nervous system. Pests scare away from the sound source as a result. This is an excellent implementation of an IoT system for pest control by Saranya et al. [49].

Another contribution of digital agriculture in the area of pest control is smart traps. Many of these traps are based on IoT. A smart trap typically uses attractants based on the specific crop and pests. These traps are strategically placed in the field. These traps can sense incoming insects and use cameras inside the trap to take pictures of the intruder. Various applications of machine vision can help us gather more knowledge after the capture. These pictures are used to compare with an existing dataset to decide if the incoming species is harmful or not. Based on the analysis the insects are then ejected or captured. A notification of the capture along with pictures of the captive can be sent to an authorized person, who can then check that specific area of his land and decide on his actions [50].

There aren't many investigations that use AI to identify pests or classify plant diseases with an emphasis on regionally important crops or insects. A deep learning algorithm based on transfer learning that can recognize the four most important Jute pest classes was proposed by Sourav et al. [51]. Jute is an excellent fiber used to make biodegradable bags, carpets, and other commonly used products. Farmers can use this sort of automatic pest detection system by the use of a smartphone app. This can save a lot of human labor. Such technologies can help to identify pest attacks at the earliest stage of the infection. Different machine-learning models for different crops are now being experimented on to further improve automatic pest detection technology [52, 53].

A smart or intelligent sprayer is another contribution of digital agriculture. These sprayers use concepts of machine vision, deep learning, and various sensors (laser scanning sensor, speed sensor) for target detection, and pesticides can be precisely applied to the desired/ targetted locations. These sprayers can help in weed management, disease treatment, pest control, etc. which can eliminate labor costs and health hazards related to pesticide spraying. Also, smart sprayers can significantly reduce the amount of pesticide usage. The amount of pesticide usage reduced by smart sprayer compared to conventional spraying technique for four different fruits according to Chen et al. [54] is described in Table 1:

| Fruit | Reduced pesticide usage by |
|---|---|
| Apple | 58.7% |
| Peach | 30.6% |
| Blueberry | 47.9% |
| Black raspberry | 52.5% |

**Table 1: Pesticide usage reduce (percentage) with smart spraying for 4 fruits**

Use of drones for smart spraying technology has also been quite successful. One such successful implementation mounted a camera to capture photos and a neural network to identify a specific pest (T. papillosa). The spraying path is then optimized by the drone. This drone can also send users valuable real-time data regarding the location and density of pests in the field through a smartphone or computer. The user can plan his actions based on this information. This specific model has been developed by Chen et al. [55]. The use of smart spraying technology can limit human interaction with harmful substances and apply pesticides over a large land far more efficiently and quickly than humans.

Applications of remote sensing have also contributed to disease and pest detection systems. Remote sensing is the process of acquiring information remotely, without physical contact or on-site inspection. Remote sensing works by collecting and processing electromagnetic radiation from the target object. Visible and near-infrared (VIS–NIR) spectroscopy sensors can help detect pests and diseases. Remote sensing systems based on thermal infrared and fluorescence can help in identifying symptoms of a disease or pest attack in the respiration and photosynthesis process of plants. There are other systems based on SAR & Lidar that are still in their development phase and can surely help in the future [56].

The intensification of food demand has encouraged farmers to use more and more pesticides for achieving higher production yields. This has adverse effects on environmental pollution and financial loss. This issue is not adequately addressed, especially in developing countries. For the financial and operational betterment of farmers and also to ensure food security around the globe smart pest control systems have proven to be a must.

### 3.3 Soil Mapping, Fertility Analysis, and Fertilization

The potential yield of a certain crop in a specific piece of land depends on a variety of factors such as; weather conditions, soil moisture, fertility, pH level, etc. Soil fertility analysis and crop yield prediction are some of the most important topics in digital agriculture. With an ever-increasing population and food demand comes continuous cropping. As a result arable land fertility has been depleting rapidly. Fertilizer usage without proper knowledge and analysis is not enough to solve the problem. Extensive analysis of soil nutrients and intelligent soil management decisions based on accurate data is necessary to cope with the increasing demand for food. Also, the decision of farming a crop on a specific piece of land needs careful consideration based on data.

Sensors for extracting information from soil samples have been extensively studied. Depending on their principles, soil sensors are categorized in mainly four classes: 1. Optical, 2. Electromagnetic/Electrical, 3. Electrochemical, 4. Mechanical. [57]

Electrical or electromagnetic sensors can be used to derive many important soil properties such as soil texture, salinity, moisture, etc.  Low cost and rapid response are the major advantages of these sensors. Optical sensors achieve effective soil mapping by the principles of spectroscopy - light reflectance, absorption, and transmittance. There has been success from researchers in measuring soil OM (organic matter) and moisture content and various soil nutrients such as CEC, pH, etc. For analysis electrochemical sensors generally requires more time than other soil sensors. These sensors are mostly based on the principle of an ion-selective electrode or an ion-selective field effect transistor (ISFET). These sensors enable high-resolution soil pH, nitrate($NO_3^-$), and potassium ($K^+$) level mapping [58]. Soil mechanical resistance can be measured by mechanical soil sensors. The provided capability of measuring soil strength provides insight into soil compaction. The amount of water and nutrients that a plant can get is limited by compacted soils because they slow down crop roots' speed of growth [58]. In addition to these four types discussed above, Acoustic and pneumatic sensors have also been tested for determining different properties of soil.

Traditional soil analysis is laboratory-dependent. When soil is removed from it is natural environment, after 24 hours the nutrient proportion in that soil sample may not be the same. Hence analysis of such samples can provide inaccurate results, which in turn can result in wrong decision-making. This is especially a challenge for remote areas with minimum transportation. Like in every other aspect of digital agriculture IoT based systems have come up with multiple renditions of soil fertility analysis. More et al proposed an IoT system based on color sensors [59]. In this system, a chemical kit is provided to the farmers to mix with the soil samples in three tubes. These three tubes are analyzed with a color sensor and data is sent to the cloud via a microcontroller. The data is then analyzed and the percentage of Nitrogen, Phosphorus, and Potassium is displayed. Use of fertilizers in proper proportion based on these data can improve crop yield and also take informed soil management decisions. Such technology also minimizes time and effort relating to sample collection and testing in soil fertility analysis.

Digital soil mapping can be defined as the process of mapping soil attributes/nutrients etc. by numerical models inferring the spatial and temporal variations of soil types and soil properties from soil observation. Such extensive mapping of soil requires a lot of data. But traditional and available soil survey data resources are sporadically available. Conventional soil surveys are also time-consuming and laborious. But recent advancements in computation power, Geographic Information System (GIS), GPS, Remote and proximal sensing technologies have made the mapping technology better. These improvements can be categorized into two parts: Input (Data collection techniques, sensors) and predictive models (Machine learning, AI, Statistical models, Geo-statistics). A framework of digital soil mapping proposed by ZHANG et al. [60] is as follows in Figure 2.

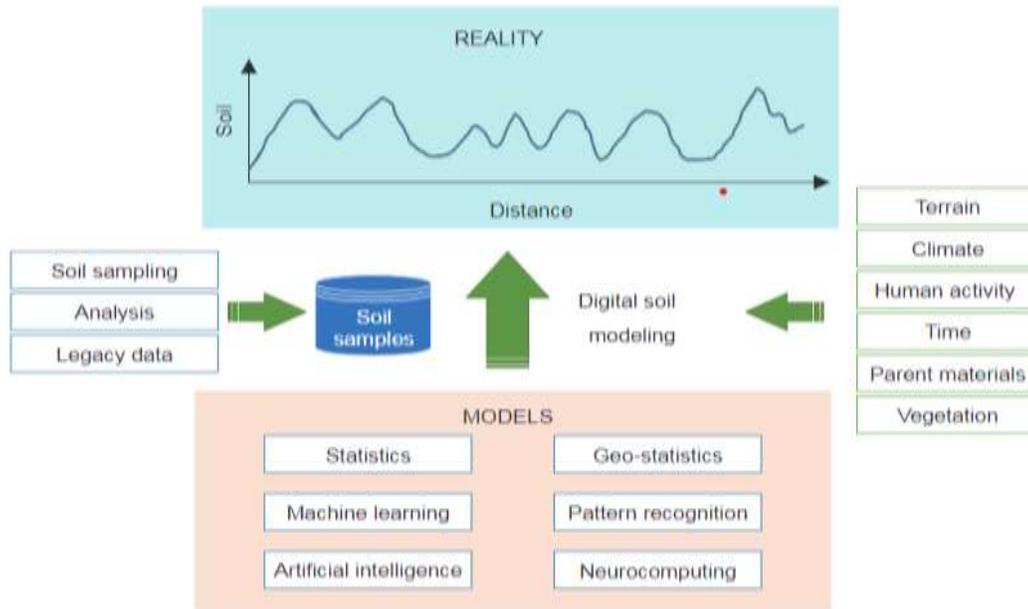

**Figure 2.** Digital Soil Mapping Framework [60]

One of the most important parts of agriculture is fertilization. Fertilization is the process of supplementing the nutrients absorbed by the latest batch of crops. It is essential for better yield. One modern idea of reshaping fertilization is variable rate fertilization or Variable rate nutrient application (VRNA). The core idea behind variable rate application is to use fertilizer, water, seed, etc. across the field, in different quantities depending on the conditions of that particular area. variable rate application is economically and environmentally efficient for obvious reasons. The application of variable rate fertilization is dependent on proper soil mapping and soil fertility analysis as soil properties vary spatially. Classifying the land in discussion into soil mapping units or management zones is an adequate approach to handling spatial variability [61, 62].

The implementation of precision fertilization can be divided into four major steps: 1. Field information collection 2. information management and decision-making 3. Execution systems [63]. Information collection is the first step when we gather data using different monitoring technologies such as GPS, Remote sensing, GIS, etc. Proper usage of such data acquisition technologies have improved the speed and accuracy of surveying and information retrieval. Farmland under strict supervision of state-of-the-art sensors can provide valuable insights regarding crop growth rate, and soil nutrient level. For decision making an array of Decision Support Systems (DSS) have been analyzed in recent years. These are interactive computer-based tools to help make a decision based on data. Some DSS examples are: Fuzzy logic, profile matching, Adaptive Neuro-Fuzzy Inference System (ANFIS), Analytic Hierarchy Process (AHP), etc. Decision Support Systems are not created to optimize decision-making, instead, they present different solutions to the decision-makers [64]. The execution systems for variable rate fertilization include automated machinery (fertilizer ejectors, Variable spraying devices, etc.) aimed at providing high fertilization accuracy.

Extensive soil analysis and smart fertilization techniques are of paramount importance for lands to sustain good crop production. Apart from the evident economic benefits these modern technologies can help preserve the ecosystem by preventing soil erosion, and infertility. Excessive use of fertilizers can be avoided by making smart decisions based on data, which in turn can prevent environmental pollution.

### 3.4 Weed management

Weed is a significant problem in agricultural ecosystems. Weed can be defined as a plant growing in an area where it is unwanted and needs management to prevent it from interfering with crop production. In general, weed has a vigorous growth rate. Losses incurred by farmers due to weed take a heavy toll on their income. Expenditure in weed management accounts for almost 33% and 22% of the total cost of cultivation, in the rainy and winter season respectively, in a survey conducted in India [65]. Weeds compete with the crop for resources including light, soil nutrients, space, etc. and affects both the quantity and quality of crop production. Some specific weeds can release substances that are toxic to the crop. This specific phenomenon is known as Allelopathy. Weeds also create problems by harboring pests and diseases. Usage of herbicides and manual removal of weeds by hand are traditional ways to prevent this undesirable vegetation. Manual removal/ hand pulling of weed is not efficient or practical for a large spread of land and may also be costly, and labor intensive. Excessive and uncontrolled use of herbicides can harm the environment as well as crops. So the introduction of more efficient techniques is required to meet ever-increasing global food demand.

The usage of agrochemicals/herbicide sprayers to control weeds has decreased human involvement a lot. But usage of regular sprayers can result in inefficient usage, as weeds grow in patches a lot of the time. Also spraying without proper analysis can create problems such as unnecessary applications of herbicides, detrimental residue in food, etc. To eliminate such problems smart spraying technologies are ideal choices. In one such iteration of a smart sprayer for weed mapping and management machine vision based on deep learning was used to distinguish between crop and weed by Partel et al. [66]. Then through the use of GPS data herbicide spraying was carried out on the identified targets only. Variable rate application can be carried out using such advanced sprayers which can significantly reduce input, and cost. Smart sprayer technology is also time efficient compared to regular methods of weed management. Continuous application of the same herbicide can cause ecological imbalance, herbicide-resistant weeds, etc [67]. Hence alternative approaches are necessary to mitigate dependency on herbicides.

For the successful implementation of smart sprayers accurate weed detection methods are imperative. Various machine learning and deep learning approaches have been analyzed. The lack of labeled data is the biggest challenge, researchers are facing at the moment. Different weather condition is also a challenge when it comes to weed detection. As plants grow randomly, leaf occlusion hinders accurate prediction of weed location.

For weed sensing or weed mapping over a large area remote sensing technology is one of the most viable options [68]. Satellite imaging can be an option for weed occurrence mapping because of its area of coverage. But image quality can be an issue in the usage of satellite images. Mainly because weed patches may not be detectable in an image with lower pixel density. Advanced cameras mounted on Unmanned aerial vehicles (UAV) can solve the problem of image quality. UAVs might not have the territorial coverage of satellites but they offer flexibility. UAVs in coordination with GPS and GIS can efficiently map instances of weed presence over a large farming land. Multispectral and hyperspectral imaging sensors can further improve this method of weed mapping. These sensors collect hundreds of images of the same spatial area in different wavelengths, which enables much better information retrival compared to regular RGB images. One such hyperspectral imaging-based weed identification system was able to successfully identify 95%, 94%, and 99% of tomatoes, black nightshade, and pigweed, respectively [69].

Recent advances in automation and robotics mean robots are trying to substitute human involvement for weeding. Many commercially manufactured options are available for weddings. LaserWeeder is one such option from a company named Carbon Robotics. The latest version of LaserWeeder specializes in large-scale row crops and involves tractor-towing for movement. It uses high-resolution cameras in accordance with deep learning to distinguish weeds from crops and uses thermal energy. They claim to have achieved an average effective weeding capacity of two acres per hour [70]. Small Robot Company created a combination of three autonomous robots to identify, eliminate weeds and sow seed. These robots are named Tom, Dick, and Harry [71]. These three robots can be used in clear soil before sowing seeds. Tom is tasked with locating weeds and Dick zaps weeds with electricity according to Tom's findings. Harry sows seeds and can record the location for future reference. Robocrop intra-row and InRow Weeder by Garford is another significant weeding solution. In their working principle, a camera is used to capture images of the crop. Depending on the position of the crop weeder discs rotate laterally to get rid of weeds. This unit can successfully detect crop plants in a row and between two rows to efficiently rotate weeder mechanisms [72]. All of the above company's applications focused on minimal usage of chemicals/ herbicides. Another company called EcoRobotix has an autonomous robot called AVO. This robot can spray very small doses of herbicides. They claim to have precision down to the centimeters while spraying [73]. Another significant feature of AVO is that it is solar-powered. Other notable commercial weeding robots include OZ, DINO by Naïo Technologies, UP by AGRIO, Tertill, etc [74, 75, 76, 77]. One important observation is all of these systems work well for row crops. So there is still a lot of room for improvement.

Environmental, operational, and financial benefits of digital agriculture in weed management are evident from our discussion. To successfully implement concepts such as Site Specific Weed Management (SSWM) and Precision Agriculture, the digital revolution has a major role to play.

**4 Conclusion**

The role of digital agriculture in transforming rural areas into smart villages is becoming increasingly important. With the rapid advancement of technology, digital agriculture is a key component of making rural areas more efficient, productive, and accessible to the public. In this study we discussed the facts that digital agriculture has the potential to revolutionize rural areas by providing access to the latest information and tools, increasing efficiency and productivity, and promoting economic growth. Digital agriculture uses technology to improve the quality of life in rural areas. This includes access to new information and tools such as precision agriculture, remote sensing, and GIS mapping. By providing farmers with access to new information and tools, digital agriculture can help them to increase their yields and reduce costs. Digital agriculture can also allow for better management of resources such as soil, water, and labor. Farmers can use digital agriculture to optimize their crop production and manage their resources more efficiently. In addition to improving the quality of life in rural areas, digital agriculture can also help to promote economic growth. By providing access to new information and tools, digital agriculture can help to increase the value of the land and increase the productivity of farmers. This can lead to an increase in agricultural output and greater economic opportunities. By increasing the value of the land, digital agriculture can also help to attract new businesses as well.

**Reference**

[1] ENRD, "'SMART VILLAGES REVITALISING RURAL SERVICES European Network for Rural Development,'" *2018*, no. 26, pp. 1–52, 2018, [Online]. Available:https://enrd.ec.europa.eu/publications/eu-rural-review-26-smart-villages-revitalising-rural-services_en

[2] J. Holmes and M. Thomas, "Introducing the Smart Villages Concept," *Int. J. Green Growth Dev.*, vol. 2, no. 2, pp. 151–154, 2015.

[3]     I. International Telecommunication Union, *Building Smart Villages: A blueprint As piloted in Niger*. 2020.

[4]     A. Martinez Juan and J. McEldowney, "Smart villages Concept, issues and prospects for EU rural areas," no. March, 2021.

[6]     E. Commission and D.-G. for A. and R. Development, *Pilot project : smart eco-social villages : final report*, no. April. 2020.

[7]     S. Stojanova, G. Lentini, P. Niederer, T. Egger, N. Cvar, and A. Kos, "Smart Villages Policies : Past , Present and Future sustainability Smart Villages Policies : Past , Present and Future," no. February, 2021, doi: 10.3390/su13041663.

[8]     "What is a Smart Village – Smart Villages." https://nkwlsmartvillages.ie/what-is-a-smart-village/ (accessed Sep. 24, 2022).

[9]     V. Zavratnik, A. Kos, and E. S. Duh, "Smart villages: Comprehensive review of initiatives and practices," *Sustain.*, vol. 10, no. 7, 2018, doi: 10.3390/su10072559.

[10]    S. Kalinowski, Ł. Komorowski, and A. Rosa, "The smart village concept. Examples from Poland," *smart village concept. Examples from Pol.*, 2022, doi: 10.53098/978-83-961048-1-6.

[11]    T. Group and S. Villages, "Collection of projects presented by TG members Working document," no. February, 2018.

[12]    M. Gascó-Hernandez, "Building a smart city: Lessons from Barcelona," *Commun. ACM*, vol. 61, no. 4, pp. 50–57, Apr. 2018, doi: 10.1145/3117800.

[13]    H. Kimanthi and P. Hebinck, "'Castle in the sky': The anomaly of the millennium villages project fixing food and markets in Sauri, western Kenya," *J. Rural Stud.*, vol. 57, no. October 2021, pp. 157–170, 2018, doi: 10.1016/j.jrurstud.2017.12.019.

[14]    S. Mitchell *et al.*, "Articles The Millennium Villages Project : a retrospective , observational , endline evaluation," *Lancet Glob. Heal.*, vol. 6, no. 5, pp. e500–e513, doi: 10.1016/S2214-109X(18)30065-2.

[15]    G. P. Green, "Handbook of Rural Development-Edward Elgar Pub (2014)," p. 358 p., 2013.

[16]    T. Van Gevelt and J. Holmes, "A Vision for Smart Villages," *Smart Villages New thingking off-grid communities Worldw.*, vol. 5, no. 5, pp. 1–6, 2015, [Online]. Available: www.e4sv.org

[17]    CORK 2.0, "A Better Life in Rural Areas Considerations (Cork 2.0 declaration)," pp. 1–5, 2016, doi: 10.2762/618522.

[18] T. V. Declaration, "The Venhorst Declaration," no. October, pp. 1–4, 2017.

[19] EC, "The Future of Food and Farming The Future of Food and Farming," pp. 1–27, 2017.

[20] European Commission, "Smart villages Bled declaration - for a Smarter Future of the Rural Areas in EU," no. April, pp. 1–2, 2018, [Online]. Available: http://pametne-vasi.info/wp-content/uploads/2018/04/Bled-declaration-for-a-Smarter-Future-of-the-Rural-Areas-in-EU.pdf

[21] S. Kalinowski, Ł. Komorowski, and A. Rosa, *The smart village concept. Examples from Poland*, no. June. 2022. doi: 10.53098/978-83-961048-1-6.

[22] S. Lanka, A. Sector, and E. Farmer, "Save and Grow Approach for Climate Resilient Agriculture," 2018.

[23] "Climate-Smart Agriculture." https://www.worldbank.org/en/topic/climate-smart-agriculture (accessed Sep. 30, 2022).

[24] *Climate-smart agriculture case studies 2021*. 2021. doi: 10.4060/cb5359en.

[25] J. Park and S. Lee, "Smart Village Projects in Korea: Rural Tourism, 6th Industrialization, and Smart Farming," *Smart Villages EU Beyond*, pp. 139–153, 2019, doi: 10.1108/978-1-78769-845-120191011.

[26] L. A. Rudwiarti, A. Pudianti, A. W. R. Emanuel, V. R. Vitasurya, and P. Hadi, "Smart tourism village, opportunity, and challenge in the disruptive era," *IOP Conf. Ser. Earth Environ. Sci.*, vol. 780, no. 1, Jun. 2021, doi: 10.1088/1755-1315/780/1/012018.

[27] R. Ciolac, T. Iancu, G. Popescu, T. Adamov, A. Feher, and S. Stanciu, "Smart Tourist Village—An Entrepreneurial Necessity for Maramures Rural Area," *Sustainability*, vol. 14, no. 14, p. 8914, 2022, doi: 10.3390/su14148914.

[28] W. Z. Li and H. Zhong, "Development of a smart tourism integration model to preserve the cultural heritage of ancient villages in Northern Guangxi," *Herit. Sci.*, vol. 10, no. 1, pp. 1–15, 2022, doi: 10.1186/s40494-022-00724-3.

[29] ENRD European Network for Rural Development, "Strategy for Inner Areas, Italy. Working document," pp. 1–4, 2018, [Online]. Available: https://enrd.ec.europa.eu/sites/enrd/files/tg_smart-villages_case-study_it.pdf

[30] ENRD, "Countryside study," pp. 1–4, 2018.

[31] D. Azevedo, "Smart Villages in the EU and Beyond Article information :," 2019.

[32] "Precision Ag Definition | International Society of Precision Agriculture." https://ispag.org/about/definition (accessed Oct. 06, 2022).

[33] A. Adesipo *et al.*, "Smart and climate-smart agricultural trends as core aspects of smart village functions," *Sensors (Switzerland)*, vol. 20, no. 21, pp. 1–22, 2020, doi: 10.3390/s20215977.

[34] World Health Organization. "Imbalances in rural primary care: a scoping literature review with an emphasis on the WHO European Region." (2018).

[35] Anyamele OD. 2009. Urban and rural differences across countries in child mortality in sub-SaharanAfrica.J. Health Care Poor Underserved20:90–98

[36] Blaauw D, Erasmus E, Pagaiya N, Tangcharoensathein V, Mullei K, et al. 2010. Policy interventionsthat attract nurses to rural areas: a multicountry discrete choice experiment.Bull. World Health Organ.88:350–56

[37] Corbett, M. (2015). Rural education: Some sociological provocations for the field. *Australian and International Journal of Rural Education*, 9-25.

[38] Daskalakis, Spyridon Nektarios, et al. "A uW backscatter-morse-leaf sensor for low-power agricultural wireless sensor networks." *IEEE Sensors Journal* 18.19 (2018): 7889-7898.

[39] Obaideen, Khaled, et al. "An overview of smart irrigation systems using IoT." *Energy Nexus* (2022): 100124.

[40] FAO. AQUASTAT: Water Uses. 2016. Available online: http://www.fao.org/nr/water/aquastat/water_use (accessed on 5 January 2019).

[41] Bwambale, Erion, Felix K. Abagale, and Geophrey K. Anornu. "Smart irrigation monitoring and control strategies for improving water use efficiency in precision agriculture: A review." Agricultural Water Management 260 (2022): 107324.

[42] Taghvaeian, Saleh, et al. "Irrigation scheduling for agriculture in the United States: The progress made and the path forward." *Transactions of the ASABE* 63.5 (2020): 1603-1618.

[43] Munir, M. Safdar, et al. "Design and implementation of an IoT system for smart energy consumption and smart irrigation in tunnel farming." *Energies* 11.12 (2018): 3427.

[44] Liao, Renkuan, et al. "Development of smart irrigation systems based on real-time soil moisture data in a greenhouse: Proof of concept." *Agricultural Water Management* 245 (2021): 106632.

[45] García, Laura, et al. "IoT-based smart irrigation systems: An overview on the recent trends on sensors and IoT systems for irrigation in precision agriculture." Sensors 20.4 (2020): 1042.

[46] Munir, M. Safdar, et al. "Intelligent and smart irrigation system using edge computing and IoT." *Complexity* 2021 (2021).

[47] Jamroen, Chaowanan, et al. "An intelligent irrigation scheduling system using low-cost wireless sensor network toward sustainable and precision agriculture." *IEEE Access* 8 (2020): 172756-172769.

[48] OERKE, E.-C. "Crop Losses to Pests." The Journal of Agricultural Science, vol. 144, no. 1, 2006, pp. 31–43., doi:10.1017/S0021859605005708.

[49] Saranya, K., et al. "IoT based pest controlling system for smart agriculture." *2019 International Conference on Communication and Electronics Systems (ICCES)*. IEEE, 2019.

[50] Figueiredo, Vitor Alexandre Campos, Samuel Mafra, and Joel Rodrigues. "A proposed iot smart trap using computer vision for sustainable pest control in coffee culture." *arXiv preprint arXiv:2004.04504* (2020).

[51] Sourav, M.S.U., Wang, H. Intelligent Identification of Jute Pests Based on Transfer Learning and Deep Convolutional Neural Networks. Neural Process Lett (2022).

[52] Zhao, Nan, et al. "Development of an automatic pest monitoring system using a deep learning model of DPeNet." *Measurement* (2022): 111970.

[53] Nguyen, Tuan T., Quoc-Tuan Vien, and Harin Sellahewa. "An efficient pest classification in smart agriculture using transfer learning." *EAI Endorsed Transactions on Industrial Networks and Intelligent Systems* 8.26 (2021): 1-8.

[54] Chen, Liming, et al. "Control of insect pests and diseases in an Ohio fruit farm with a laser-guided intelligent sprayer." *HortTechnology* 30.2 (2020): 168-175.

[55] Chen, Ching-Ju, et al. "Identification of fruit tree pests with deep learning on embedded drone to achieve accurate pesticide spraying." *IEEE Access* 9 (2021): 21986-21997.

[56] Zhang, Jingcheng, et al. "Monitoring plant diseases and pests through remote sensing technology: A review." *Computers and Electronics in Agriculture* 165 (2019): 104943.

[57] Molin, José Paulo, and Tiago Rodrigues Tavares. "Sensor systems for mapping soil fertility attributes: Challenges, advances, and perspectives in brazilian tropical soils." *Engenharia Agrícola* 39 (2019): 126-147.

[58] Adamchuk, Viacheslav I., et al. "On-the-go soil sensors for precision agriculture." *Computers and electronics in agriculture* 44.1 (2004): 71-91.

[59] More, Abhishek, et al. "Soil Analysis Using Iot." *2nd International Conference on Advances in Science & Technology (ICAST)*. 2019.

[60] ZHANG, Gan-lin, L. I. U. Feng, and Xiao-dong SONG. "Recent progress and future prospect of digital soil mapping: A review." *Journal of Integrative Agriculture* 16.12 (2017): 2871-2885.

[61] Iticha, Birhanu, and Chalsissa Takele. "Digital soil mapping for site-specific management of soils." *Geoderma* 351 (2019): 85-91.

[62] Cullum, Carola, et al. "Landscape archetypes for ecological classification and mapping: the virtue of vagueness." *Progress in physical geography* 41.1 (2017): 95-123.

[63] Lu, Yue, et al. "Precision Fertilization and Irrigation: Progress and Applications." AgriEngineering 4.3 (2022): 626-655.

[64] Indahingwati, Asmara, et al. "Comparison analysis of TOPSIS and fuzzy logic methods on fertilizer selection." *Int. J. Eng. Technol* 7.2.3 (2018): 109-114.

[65] Yaduraju, N. T., and J. S. Mishra. "Smart weed management: A small step towards doubling farmers' income." (2018): 1-5.

[66] Partel, Victor, Sri Charan Kakarla, and Yiannis Ampatzidis. "Development and evaluation of a low-cost and smart technology for precision weed management utilizing artificial intelligence." Computers and electronics in agriculture 157 (2019): 339-350.

[67] Gnanavel, I. "Eco-friendly weed control options for sustainable agriculture." Science International (Dubai) 3.2 (2015): 37-47.

[68] Monteiro, António, and Sérgio Santos. "Sustainable approach to weed management: The role of precision weed management." Agronomy 12.1 (2022): 118.

[69] Zhang, Y., et al. "Precision automated weed control using hyperspectral vision identification and heated oil." 2009 Reno, Nevada, June 21-June 24, 2009. American Society of Agricultural and Biological Engineers, 2009.

[70] "Carbon Robotics Unveils New LaserWeeder with 30 Lasers to Autonomously Eradicate Weeds." Carbon Robotics, Carbon Robotics, 7 Feb. 2022, https://carbonrobotics.com/news/carbon-robotics-unveils-new-laserweeder-with-30-lasers-to-autonomously-eradicate-weeds.

[71] Fund, Environmental Defense. "Meet Tom, Dick, and Harry." Medium, The Fourth Wave, 22 Feb. 2019, https://medium.com/the-fourth-wave/meet-tom-dick-and-harry-17dc6c417f9a.

[72] "Robocrop INROW Weeder: Remove Inter Row Weeds." Garford Farm Machinery, 19 July 2018, https://garford.com/products/robocrop-inrow-weeder/.

[73]  "Our Vision for the Future: Autonomous Weeding (in Development)Avo." Ecorobotix, https://ecorobotix.com/en/avo/.

[74]  "Robot Polyvalent Autonome De Semis Et De Désherbage Oz." Naïo Technologies, https://www.naio-technologies.com/oz/.

[75]  "Dino, Le Robot De Désherbage Pour Les Légumes En Planches." Naïo Technologies, https://www.naio-technologies.com/dino/.

[76]  *Autonomous. simple. affordable.* AIGRO. (n.d.). Retrieved January 7, 2023, from https://www.aigro.nl/

[77]  "Garden Happy - It's Easy with Tertill!" Tertill, https://tertill.com/.

[78]  Green, F. (2013). Demanding work. In *Demanding Work*. Princeton University Press.

[79]  Getahun, H., Matteelli, A., Abubakar, I., Aziz, M. A., Baddeley, A., Barreira, D., ... & Raviglione, M. (2015). Management of latent Mycobacterium tuberculosis infection: WHO guidelines for low tuberculosis burden countries. *European Respiratory Journal*, *46*(6), 1563-1576.